\def\year{2019}\relax
\documentclass[letterpaper]{article} 
\usepackage{aaai19}  
\usepackage{times}  
\usepackage{helvet} 
\usepackage{courier}  
\usepackage[hyphens]{url}  
\usepackage{graphicx} 
\usepackage{graphicx}  
\usepackage[caption=false]{subfig}
\usepackage[ampersand]{easylist}
\usepackage[dvipsnames]{xcolor}
\usepackage{booktabs}

\title{Understanding the Impact of Text Highlighting in Crowdsourcing Tasks}
\author{
Jorge Ram\'irez,\textsuperscript{\rm 1} 
Marcos Baez, \textsuperscript{\rm 1} Fabio Casati, \textsuperscript{\rm 1} 
Boualem Benatallah \textsuperscript{\rm 2}\\
\textsuperscript{\rm 1}University of Trento, Italy\\
\textsuperscript{\rm 2} University of New South Wales, Australia\\
\{jorge.ramirezmedina, marcos.baez, fabio.casati\}@unitn.it,
boualem@cse.unsw.edu.au
}

\begin{document}

\maketitle

\begin{abstract}
    Text classification is one of the most common goals of machine learning (ML) projects, and also one of the most frequent human intelligence tasks in crowdsourcing platforms.
    ML has mixed success in such tasks depending on the nature of the problem, while crowd-based classification has proven to be surprisingly effective, but can be expensive. 
    Recently,  hybrid text classification algorithms, combining human computation and machine learning, have been proposed to improve accuracy and reduce costs. One way to do so is to have ML highlight or emphasize portions of text that it believes to be more relevant to the decision. Humans can then rely only on this text or read the entire text if the highlighted information is insufficient.
    In this paper, 
    we investigate if and under what conditions highlighting selected parts of the text can (or cannot) improve classification cost and/or accuracy, and in general how it affects the process and outcome of the human intelligence tasks. 
    We study this through a series of crowdsourcing experiments running over different datasets 
    and with task designs imposing different cognitive demands. Our findings suggest that highlighting is effective in reducing classification effort but does not improve accuracy - and in fact, low-quality highlighting can decrease it.
\end{abstract}

\section{Introduction}


Text classification is one of the most fundamental problems of machine learning (ML) projects \cite{DBLP:books/sp/mining2012/AggarwalZ12b}, and also one of the most frequent human intelligence tasks in crowdsourcing platforms. 
It also occurs naturally in many activities we are faced in our work as scientists, such as identifying if a paper is relevant to a research topic \cite{DBLP:journals/jamia/WallaceNMCST17}.

While ML has done impressive progress in some domains, it is still unable to accurately classify in many complex contexts. In the latter case we can resort to crowdsourcing, but this can be expensive especially when the problem is challenging or the text is long.


Recently,  \textit{hybrid} text classification algorithms, combining human computation and machine learning, have been proposed to improve accuracy and reduce costs. 
These techniques capitalize on the strength of humans and of machine classifiers to solve difficult tasks \cite{KrivosheevCSCW2018,DBLP:conf/nips/GomesWKP11,DBLP:conf/aamas/KamarHH12,DBLP:conf/cscw/ChengB15}. 

One way to capitalize on these complementary strengths is to have ML highlight or emphasize portions of text that it believes to be more relevant to the decision. Humans can then rely only on this text or read the entire text if the highlighted information is insufficient.
Indeed, researchers in information management and psychology have shown that text highlighting can improve the reading time of humans \cite{DBLP:journals/iam/WuY03}.
However, it can also be harmful when it is inappropriate or not relevant \cite{Gier2009HarmfulEO}. 
%


Previous research has explored the benefits of highlighting in: supporting workers in digitization tasks by highlighting target fields \cite{alagarai2014cognitively}, recommending text excerpts to facilitate the job of text annotators \cite{Wilson2016WWW}, requesting highlights as evidence to support judgments \cite{schaekermann2018resolvable}, and as a tool to explain the output ML models \cite{DBLP:conf/naacl/Nguyen18}. 

In this paper, we study if and under what conditions highlighting excerpts from the text can (or cannot) improve text classification cost and/or accuracy, and in general how it affects the process and outcome of the human intelligence tasks.
This is important both because highlighting can  not only be a task in a two-step crowd classification procedure (highlight, then classify) but, perhaps most importantly, can also be used in hybrid classification processes where text summarization algorithms identify relevant portions of a text, thereby simplifying the subsequent (human) classification task.
We do this through a series of crowdsourcing experiments running over different datasets with varying classification difficulty and document length, and with task designs imposing different cognitive demands. 
%

Specifically, we make the following contributions:
\begin{itemize}
    \item We are, to the best of our knowledge, the first to systematically study the effect of text highlighting in human computation, identifying the quality requirements that algorithms for text highlighting should possess to help with text classification and  estimating the potential impact of good (and bad) highlighting.
    \item We uncover the potential of aggregating highlighting by multiple, independent annotators (or algorithms) showing that aggregation is practical and useful, somewhat analogously to what happens in a crowdsourced classification where we aggregate multiple votes on items.
    \item We discuss interesting and perhaps unexpected effects of highlighting, important to make them effective, such as giving time to  workers to get used to working with highlights.
    \item We contribute an annotated dataset for researchers who want to study the problem. 
\end{itemize}





\section{Related work}

Highlighting is a common tool used to mark relevant sections in text \cite{DBLP:journals/tvcg/StrobeltOKSP16}. The act of identifying what is important to highlight in a text have been shown useful for learning \cite{Craik1972}. 
Fowler and Barker \cite{Fowler1974} have shown that students had better recall of highlighted passages in a document in comparison to non-highlighted portions, after reading a document with preexisting highlighting. However, when the preexisting highlighting is inappropriate (the highlighted portions are not relevant to the content of the document), Gier et al. showed \cite{Gier2009HarmfulEO} that this could impair the reading comprehension. Besides understanding, studies have shown that highlighting could reduce the cognitive load, that is, the reading time \cite{DBLP:journals/iam/WuY03}.

In crowdsourcing, researchers have used highlighting to facilitate the job of workers. 
Alagarai et al. \cite{alagarai2014cognitively} explored different variation of a form digitation task, showing that highlighting of the target fields improved the accuracy of workers. 
Wilson and colleagues \cite{Wilson2016WWW} studied the feasibility of crowdsourcing for annotating privacy policies and how automatic highlighting of relevant paragraphs can support annotators. They showed that highlighting reduces task completion time without hurting nor improving the accuracy of the annotators. 
Besides helping workers, highlighting have also been used to ask the crowd for evidence that support their judgement \cite{schaekermann2018resolvable,DBLP:conf/hcomp/McDonnellLKE16}.

In the context of interpretability of machine learning models, text highlighting have been used to present machine-generated explanations (relevant words) to humans for evaluation. In these settings,  Nguyen \cite{DBLP:conf/naacl/Nguyen18} asked workers on AMT to guess the output of the model based on the text and the highlighted explanation, to determine how automatic evaluation compares to the human-level evaluation of explanations.



Researchers have shown the feasibility of non-expert annotations for NLP tasks \cite{DBLP:conf/emnlp/SnowOJN08}, and the above works have shed some lights on the potential of highlighting as a tool for assisting workers. However, no study discusses the effects of highlighting in crowd classification, considering the quality and quantity of the highlighted text, and the behavior of workers on documents with varying difficulty.
This is central to any study as it indicates how ``good" highlighting needs to be to provide value to crowd classification.

\section{Research Questions}


%

The problem of hybrid classification via text highlighting has two sides: i) obtaining the highlighting and ii) using it in crowdsourcing tasks.
In this paper we focus on the latter problem, that of classifying using highlighted text support. 
The first problem is relevant only in terms of obtaining a rich and diverse dataset of highlighted text, that as we will see presents challenges in itself.

We set to study the impact of highlighting under different metrics, all important to crowdsourcing: the classification \textbf{accuracy} and the \textbf{decision time} (time spent on the task which, if we set the pay rate based on time, is directly related to costs). 
In analysing the impact of highlighting, we focus particularly on the following research questions: 

\begin{easylist}
\ListProperties(Hide=100, Hang=true,Style*=--~)
& \textbf{RQ1}. Does  highlighting increase worker accuracy?
Specifically, we consider three dimensions of the problem when assessing impact of highlighting: 
i) The \textit{quality of highlighting}, meaning, whether the emphasised texts actually facilitates the classification tasks, is neutral, or possibly even hurts it, ii) the \textit{difficulty} of the classification task, and iii) the \textit{length} of the document and the proportion of highlighted text. 
& \textbf{RQ2}. Does  highlighting reduce decision time? And again, how is decision time impacted based on quality, difficulty and length?

\end{easylist}


Considering these factors is important because it helps us understand \textit{how good} highlighting needs to be in order to be useful, thereby setting the bar for human computation or ML algorithms obtaining such highlighting. It also tells us for which kinds of tasks the impact may be more or less significant.

%


\section{Crowdsourcing and Generating Highlights}
As basis for our study of highlighting effectiveness, we obtained and assessed highlights for three datasets with different properties in terms of document length and classification accuracy, used in prior art  \cite{KrivosheevCSCW2018}. 
We obtain highlights from both humans and algorithms. 
Crowdsourced highlights allow us to obtain a wide set of highlights and highlighting patterns (e.g., individual words, full sentences) and of highlighting quality for the same text. Machine highlights, obtained via  state of the art algorithms, help us assess the effectiveness of the hybrid ``highlight then classify" approach to text classification that can be achieved today, as well as enabling us to assess machine highlighting quality with respect to the downstream task of efficient and accurate classification.
Therefore, our focus here is not to improve ML algorithms but to assess how they perform.

\noindent\textit{Systematic Literature Review (SLR)}. This dataset contains a list of 900+ abstracts 
annotated by experts according to their relevance to an SLR.
The dataset defines two relevance questions (filters): \textit{1. SLR-OA}: \textit{Does the paper describe a study that involves older adults (60+)?\footnote{60 is a commonly used age limit in scientific studies}}, and \textit{2. SLR-Tech}: \textit{Does the paper describe a study that involves technology for online social interactions?} 
%
We considered each filter separately and created two datasets of 135 and 150 papers, respectively.
The papers were randomly selected but controlled for the abstract length. We first excluded a long tail of outlier abstracts of length over 4000 characters, divided the remainder in three buckets of equal number of abstracts (the dividing points turned out to be 1050 and 2150 characters) and sampled an equal number of abstracts from each bucket.

For \textit{SLR-OA} we also balanced the number of papers that described the population age explicitly vs those that refer to ``older adults" or synonyms. 
We do so as we suspect (as it turned out) that this can impact worker behavior and performance.
The distribution of the ground truth labels for \textit{SLR-OA} is 41.5\% no, 54.1\% yes, 4.4\% maybe; and for \textit{SLR-Tech} is 56\% no, 40\% yes, 4\% maybe.

\smallskip

\noindent\textit{Amazon Reviews}. The dataset contains reviews about products sold on Amazon. It includes 100k items annotated with ground truth on two relevance questions, including \textit{Is this review written on a book?}. 
We selected 400 reviews randomly (50\% about books and 50\% about other products), focusing only on short (200 reviews with $< 1050$ characters, but as long as at least the shortest SLR abstract that has 625 characters to make a fair comparison) and long reviews (200 reviews with $>2159$ characters). 


\smallskip

\noindent\textbf{Crowd-generated highlights.}
\label{section:rationales} 
To test the effects of highlighting of different quality in a controlled fashion, we ran a series of crowdsourcing tasks that requested users to classify items and highlight the reasons supporting their judgment.
We do not discuss this task further as obtaining highlighting is not the focus of this paper, but the interested reader can see the task description and results in the supplementary material\footnote{Material available at \textit{\url{https://tinyurl.com/hcomp19-hl}}}. 
We collected 3-7 
highlighted excerpts
per document and filter, totaling $2722$ highlights ($610$ for \textit{SLR-OA}, $616$ for \textit{SLR-Tech}, and $1496$ for \textit{Amazon}).


Two researchers assessed the quality of the highlighting provided by workers according to the following coding scheme:
\textit{bad}: the rationale could potentially lead a worker to make a wrong decision;
\textit{neutral}: it does not provide information to make a decision;
\textit{suboptimal}: it could potentially help but there are other fragments that are more suitable;
\textit{good}, it holds enough information that could help a worker in making the right decision. 

The procedure for coding the quality of highlights involved both coders going over 20\% of each dataset for tuning specific criteria, followed by independent coding on random splits of each dataset. Disagreements were down to a minimum and within the same usefulness class (mixing bad/neutral or good/suboptimal highlighting), the resulting Cohen's Kappa was $0.87$ for SLR-OA, $0.72$ for SLR-Tech and $0.66$ for Amazon.

\noindent\textbf{Machine-generated highlights.}\label{section:machine-highlights} We generate highlights based on two approaches: state of the art algorithms for \textit{extractive summarization} \cite{DBLP:journals/corr/BertSum,Refresh2018}, which are independent of the specific question being asked, and \textit{question-specific} highlighting \cite{devlin2018bert}. 
For the first approach, we selected \textit{BertSum} \cite{DBLP:journals/corr/BertSum} and \textit{Refresh} \cite{Refresh2018}. 
These  are recent algorithms for extractive summarization, where \textit{BertSum} produces state-of-the-art results on a commonly used dataset.
For \textit{BertSum} we followed the training procedure with the indicated dataset, and for \textit{Refresh} we used the available pre-trained model. 

%



Leveraging a ``generic" extractive summarization algorithm might give useful summaries but would not however be a fair comparison with crowd highlighting and probably not efficient for question-specific classification. We, therefore, chose to generate question-specific highlights by borrowing Q\&A algorithms that provide answers as a subset of a text (e.g., the answer is a sentence or paragraph from a Wikipedia page that the algorithm believes to contain the information necessary to answer the question).
In other words, we use Q\&A ML to obtain the \textit{rationale} that can support an answer~\cite{COQA}, but not the answer per se which is left to the crowd, who may or may not make use of the rationale as a guide or as a way to help determine the answer more quickly.

Specifically, we leverage a fine-tuned version of BERT \cite{devlin2018bert} for question answering, or \textit{Bert-QA} for brevity.
For \textit{Bert-QA}, we used the BERT-Base (uncased) pre-trained model and followed the fine-tuning procedure on the SQuAD dataset as indicated in the BERT paper.

Notice that the kinds of tasks we aim at covering include challenging tasks (such as SLR screening) requiring very high accuracy (SLR experts achieve over 0.96 accuracy, and the same is required -- and can be obtained -- by the crowd in this domain \cite{Nguyen2015,Mortensen2016crowd,KrivosheevCSCW2018}).
Today these are outside what machines can achieve when giving direct answers\footnote{see, e.g., https://rajpurkar.github.io/SQuAD-explorer/ and https://stanfordnlp.github.io/coqa/}. 


\section{Experiment Design}

\begin{figure}[h]
\centering
  \includegraphics[width=1\columnwidth]{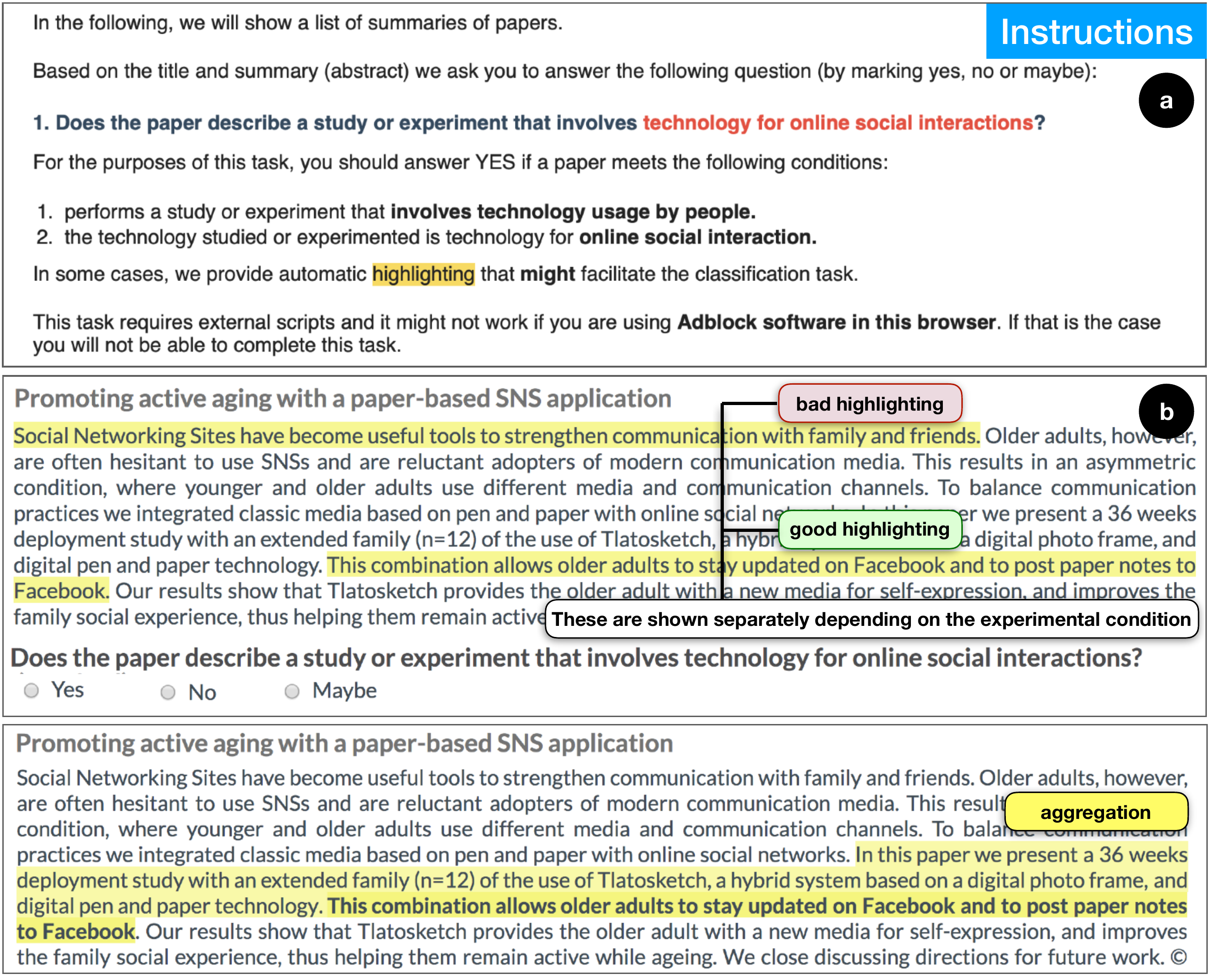}
  \caption{Task design, and example highlighting. }~\label{fig:task-design}
\end{figure}

We now have datasets with items of varying length and difficulty and with highlighting of different quality, corresponding to different control conditions.
The basic task design to assess impact is inspired by basic screening task designs \cite{CrowdRev2018,KrivosheevHcomp2017}, that have been modified to incorporate  highlighting. The task is shown in Fig \ref{fig:task-design} for SLR-Tech and is analogous for the other datasets.
Workers are presented with the text to classify, with some parts highlighted, and we mentioned that highlighting  \textit{might} (with emphasis) facilitate the classification. 

The tasks were designed and run in Figure Eight (F8)%
\footnote{Figure Eight https://figure-eight.com}. 
This platform organises the items in a task in \textit{pages}, where the first page acts as a test page (contains gold items only). Subsequent pages include a hidden test question to control for workers' accuracy. 

In the study we aim at observing the workers' \textit{accuracy}, the \textit{time to decision}, and the \textit{retention} (how many pages a worker processes before deciding to quit) as key metrics.
Notice that retention is important as dropouts make the task slower and more expensive (if a worker completes the initial tests, we are charged for the cost of test items as well, which means that we waste money if the worker abandons shortly after)\cite{DBLP:conf/wsdm/HanRGSCMD19}.


Given this setting, we run three main rounds of experiments with the following configurations.

\smallskip

\noindent\textbf{Experiment 1}, on the effects of highlighting of varying quality. 
Specifically, here we generate \textit{six} conditions: four of them contain different proportion of abstracts or reviews with ``useful" highlighting (the \textit{good} and \textit{suboptimal} highlightings were considered as \textit{useful}, while the \textit{neutral} and \textit{bad} as \textit{not useful}).
We create four conditions with 0\%, \textit{33\%}, \textit{66\%}, and \textit{100\%} useful highlighting.
Notice that the percentage refers to documents: for example, in the \textit{66\%} condition two out of three documents have only useful highlighting, while the third has non-useful ones.
The purpose of this experimental design is to assess behaviors in situations where crowd worker could consistently trust or mistrust the highlights, as well as cases where the quality is mixed.
During the qualitative assessment, the researchers generated the missing highlighting of the items (papers or reviews) with unbalanced highlights, those having only useful or not useful highlights.

%
In addition, we create an \textit{aggregation condition} that fuses, for each item, all the highlighting obtained on that item. The aggregation strategy computes a score for each word in a text as the total number of highlighting that cover the word divided by the number of workers that produced these highlighting.
With this score, the aggregation condition places more emphasis on words highlighted more often. If the score of a word is greater or equal than $0.33$, then the word is highlighted, and the opacity value is equal to the score. If the score is at least one standard deviation away from the mean, then the word is boldfaced\footnote{During our highlighting collection experiments, we developed a visual tool to evaluate the aggregation strategy and determine the values that we end up using for opacity and boldface.}.
%
An example of how aggregated highlighting looks like can be seen in Fig \ref{fig:task-design}.
Finally, we add a baseline condition where items have no highlighting. 


%
We followed a between-subject design to assign workers to one of the  0\%, 33\%, 66\%, 100\%, aggregation, and baseline condition.
We defined that a worker could give a maximum of 18 judgments divided into 6 pages of 3 items each (6$\times$3 design), and we set the accuracy threshold to be 76\% and 100\% for the SLR and Amazon datasets respectively. We set the payment to \$0.05 for the SLR datasets and \$0.02 for Amazon, aiming at a rate of 10USD/hour.
We repeatedly ran the tasks over five weeks where each lasted between 2 to 5 days, collecting votes from F8's middle tier contributors.


\smallskip

\noindent\textbf{Experiment 2} focused on the impact of highlighting on difficult and demanding tasks. We followed the same experimental setup as in Experiment 1, but modified the tasks to impose higher cognitive demands on the worker. We focused only on long documents and on the dataset with lower accuracy (\textit{SLR-Tech}), and implemented two task designs: \textit{Tech6$\times$6}, featuring longer pages with 6 documents instead of 3, while maintaining the same number of pages; and \textit{Tech12$\times$3}, featuring a longer task with 12 pages, but keeping page size.
We tested a condition with 83\% quality highlighting (based on the promising range identified in the first experiment) against the baseline. 
We paid \$0.05 per item. 

\smallskip

\noindent\textbf{Experiment 3} focused on determining if the same relationships between quality of highlights and performance are observed in scenarios that rely on automatic highlighting. This experiment relies on six experimental conditions:  three corresponding to the automatic highlight generation with \textit{BertSum}, \textit{Refresh} and \textit{Bert-QA}, an aggregation of the output of three algorithms (\textit{Aggregation-ML}), a condition with only high-quality highlighting from the automatic approaches \textit{(100\%ML)}, and a \textit{baseline} without highlighting -- the last two to provide a reference point for comparison.
The highlight of the extractive summarization algorithms (Bertsum and Refresh) is produced by taking the top ranked sentence from the resulting summary. An additional pilot is also run considering the top three sentences as the resulting highlight, so as to assess the impact of longer highlighted text.  

We followed a between-subject design to assign workers to one of the six experimental classification support conditions, and relied on the same task design (6$\times$3), datasets (\textit{SLR-OA}, \textit{SLR-Tech}, \textit{Amazon}), budget constraints and process as in Experiment 1.  

\smallskip

\begin{figure}[h]
\centering
  \includegraphics[width=.9\columnwidth]{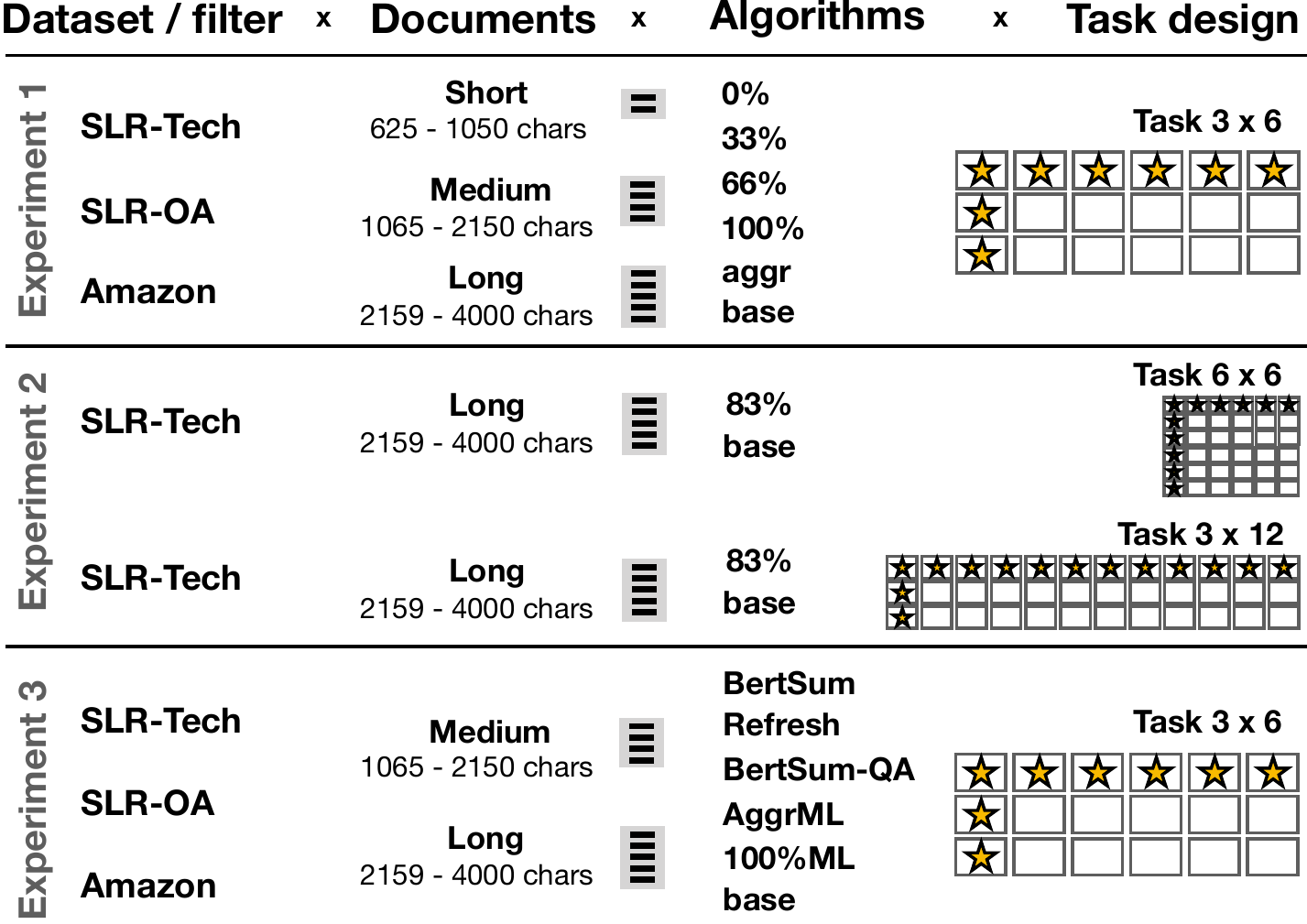}
  \caption{Experimental conditions}~\label{fig:experiment}
\end{figure}

Figure \ref{fig:experiment} shows the experimental conditions. Each task is a factorial combination of dataset, document length and design. 
An external service controlled the random assignment of workers to the conditions. 
This allowed us to run all the conditions and baseline in parallel and reduce potential noise in the results, due to the same worker taking part in multiple tasks (something we experienced in the many preparation experiments we did, and caused us to waste some of our budget). 
Specifically, we implemented an external server that keeps track of the number of workers in each condition and uses this information when a new worker arrives to perform a random assignment among the conditions with the fewer number of workers (for balancing the assignment). F8 allows adding custom JavaScript code to the task interface that runs on every page load on the workers' browsers. 
We added code to call our external server to i) determine the condition for the worker (or retrieve a previous assignment), ii) obtain what parts to highlight for each item in the current page (unless the condition is baseline), iii) compute the decision time metric. To calculate decision time, we captured each time a worker clicks on one of the possible answers and compute the difference between the first and last values stored for each of the items\footnote{We captured the page load time and used it as the starting point for computing decision time of the first item of the page}.


To avoid workers judging items of different sizes (e.g., mixing short and long abstracts in a page) we split items in the dataset along this dimension and ran separate jobs for each size bucket respectively.

During our early pilot studies, we found that most workers came from a handful of countries. 
So to avoid this potential bias, we defined three geographical buckets where the head member of each bucket was one of the top three countries identified in our pilots. We ran our experiments at three different time slots (morning, afternoon and night) to orchestrate the assignment of geographical buckets to size buckets so that at any given time slot one group of countries work on one size bucket. We swapped this assignment of countries to size buckets at each time slot to make sure that one size bucket (short abstracts, for example) gets contributions from all of our target countries. This plan for running the jobs in F8 allowed us to block workers, during a particular time frame, from jumping between jobs after they finish, that is, workers that complete judging short abstracts and then continue with the long abstracts bucket (which would bias and introduce a correlation in the results).

\section{Results}
\subsection{Experiment 1: Impact of highlighting quality}

We collected a total of 14085 judgments from 1337 workers. Table \ref{table:workers} shows the distribution of these values considering the datasets. The number of workers was balanced among the experimental conditions.


\begin{table}[h]
\centering
\resizebox{0.7\columnwidth}{!}{%
\begin{tabular}{|l|l|l|}
\hline
\textbf{Dataset} & \textbf{\#judgments} & \textbf{\#workers} \\ \hline
SLR-OA  & 3327 & 424 \\ \hline
SLR-Tech & 4014 & 464 \\ \hline
Amazon & 6744 & 449 \\ \hline
\end{tabular}%
}
\caption{Distribution of workers and judgements per dataset}
\label{table:workers}
\end{table}


\subsubsection{Worker accuracy}

\begin{figure}[h]
  \centering    
  \includegraphics[width=.9\columnwidth]{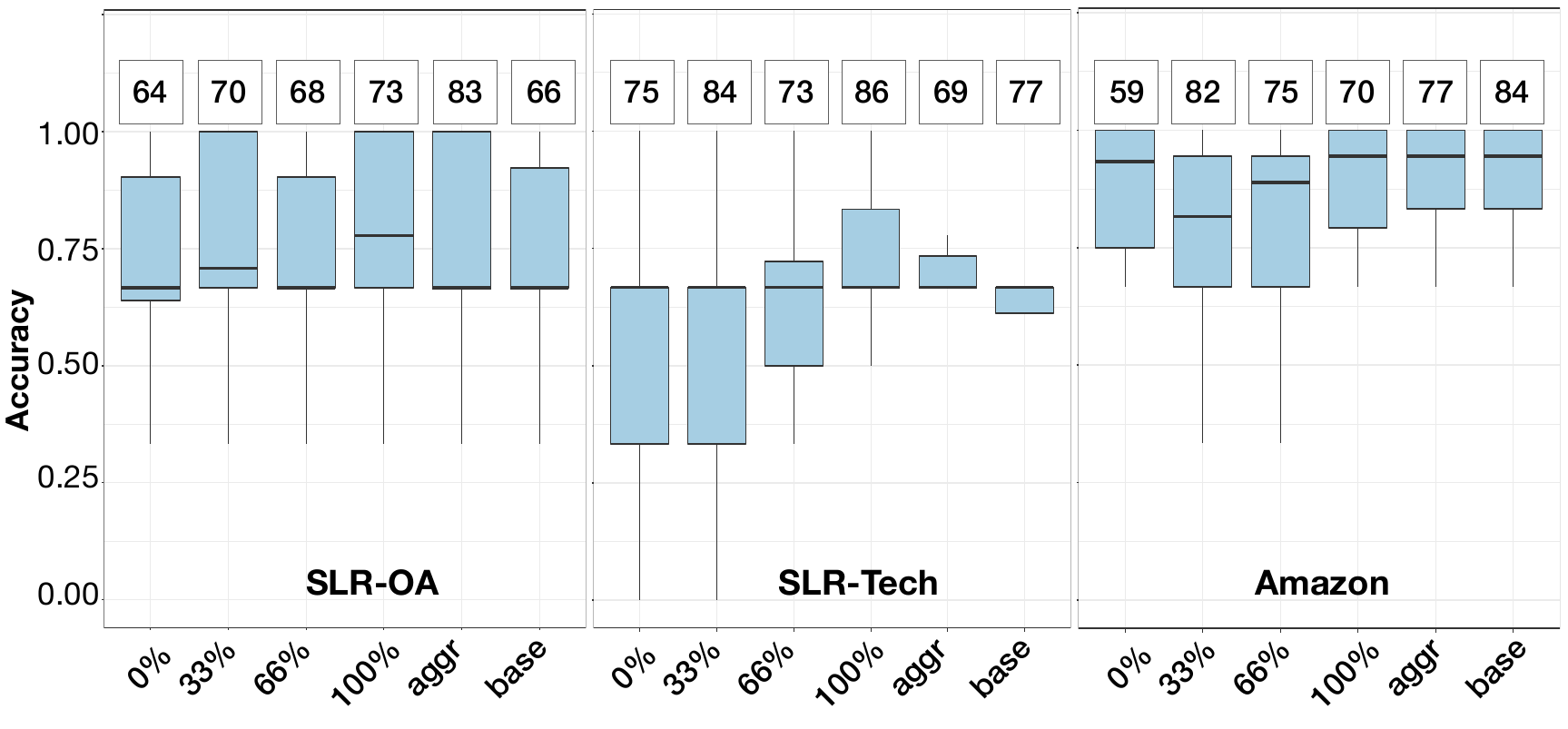}
  \caption{Worker accuracy boxplot (the top row shows the number of items in the condition).}~\label{fig:accuracy}
\end{figure}

The median accuracy of the workers in the baseline conditions was $0.67$ for \textit{SLR-OA} and \textit{SLR-Tech}, and, as expected, much higher for \textit{Amazon} ($0.94$). 
When comparing to the conditions with highlighting (see Fig. \ref{fig:accuracy}), we can see that the workers in the \textit{100\%} condition featured the same or better median accuracy (\textit{SLR-OA}: $0.78$, \textit{SLR-Tech}: $0.67$, \textit{Amazon}: $0.94$) than all the other conditions. 
%

A Kruskal-Wallis rank sum tests showed no significant difference for \textit{SLR-OA} ($H(5)=4.30$, $p=.51$), despite the trend in favor of the conditions with higher quality highlighting.
%
%
In contrast, the results for \textit{SLR-Tech} did show a statistically significant difference between the conditions ($H(5)=12.74$, $p=.03$), but with the test of multiple comparisons \cite{DunnTest1964} (using Benjamini-Hochberg adjustment) indicating a significant difference only between the extremes \textit{100\%} and \textit{0\%} in favor of the former. 
In the \textit{Amazon} dataset we also observed a statistically significant difference 
($H(5)=21.76$, $p<.001$), with the test of multiple comparisons showing the difference to be significant between  \textit{33\%} and all the others conditions, and between \textit{66\%} and \textit{100\%} -- these differences in detriment of the conditions with lower quality. 

The above tell us that despite the trend in favor of the conditions with higher quality highlighting, and in particular the \textit{100\%}, \textit{the highlighting support did not improve over the baseline. Instead, we have seen the opposite effect: bad highlighting can hurt accuracy.}

\subsubsection{Decision time}
The median decision time in the baseline conditions was $12.75s$ for \textit{SLR-OA}, $32.52s$ for \textit{SLR-Tech} and $15.62s$ for \textit{Amazon}. 
Deciding whether an abstract is related to older adults required less effort than for \textit{SLR-Tech}, we believe because  the nature of former was more suitable for screening for keywords and age (e.g., ``older adults", ``aged 60 and older"). 
Surprisingly, workers took more time in screening \textit{Amazon} reviews - a fairly easy task -  than screening abstracts with the \textit{SLR-OA} dataset. 



\begin{figure}[h]
  \centering    
  \includegraphics[width=.9\columnwidth]{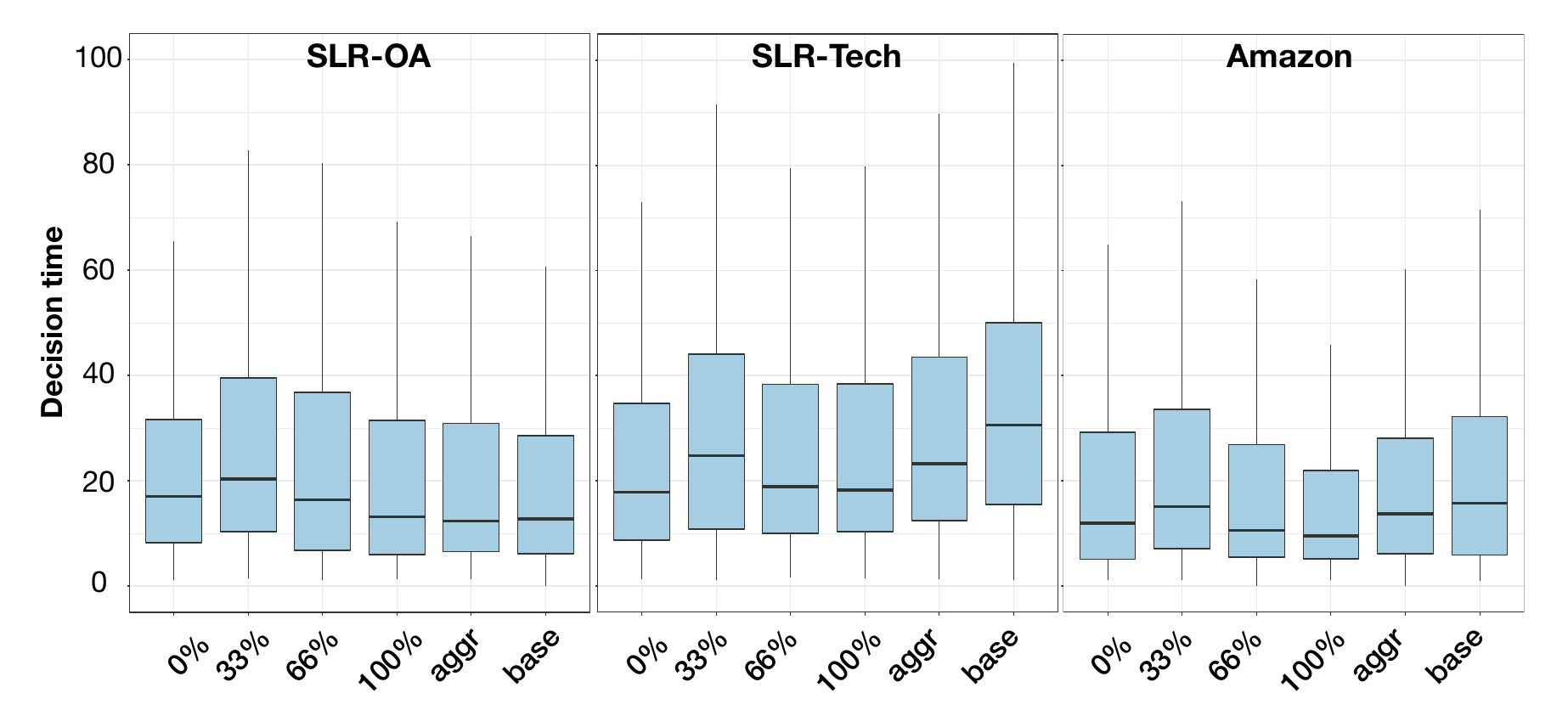}
  \caption{Decision time per condition.}~\label{fig:reading}
\end{figure}

In comparison, the best performance for the highlighting conditions improved on the baseline in all the filters  (\textit{aggr}=$12.41s$ for \textit{SLR-OA}; \textit{0\%}=$18.51s$ for \textit{SLR-tech};  \textit{100\%}=$9.45s$ for \textit{Amazon}). The general trend, as shown in Fig. \ref{fig:reading}, is that of conditions with higher-quality highlighting resulting in lower decision time, except for the curious case of \textit{0\%}, where workers achieved a performance not only better or at par with the baseline, but also with the conditions with mixed quality highlighting when considering all filters. 
We attribute this behavior to workers learning of the highlighting support not being useful (or being deceitful), which might have led to them dismissing the highlighted text and redirecting their attention to other parts of the document --- thus having a similar effect as in the \textit{100\%} condition.

Kruskal-Wallis rank sum tests show 
statistically significant difference between the conditions for all datasets ($H(5)=43.78$, $p<.001$ for \textit{SLR-OA}, $H(5)=40.31$, $p<.001$ for \textit{SLR-Tech}, and $H(5)=50.52$, $p<.001$ for \textit{Amazon}).
Multiple comparison tests show that the \textit{100\%} condition has significantly faster decision times with respect to the baseline  for \textit{SLR-Tech} and  \textit{Amazon} and it significantly outperforms all other highlighting conditions (except for \textit{SLR-Tech} where nearly every condition significantly outperforms the baseline).
The test also confirms the curious effect of the \textit{0\%} condition outperforming the \textit{33\%} one.
The detailed test results are available in the supplementary material.





Aside from the \textit{SLR-OA} dataset, the above results indicate that good-quality highlights give an advantage to workers, reducing the time to judge. The benefit is pronounced in the 66\% to 100\% range, while the worst performance can be expected when mixing good highlighting with a majority of bad highlighting. This situation has proven to harm decision time more than having all documents with low quality highlighting. 



\subsection{Experiment 2: Impact in demanding tasks}
In this experiment, we focused on understanding the impact of highlighting in situations of higher cognitive demand. 

\smallskip
We collected 2481 judgements in total from 255 workers. Of these, 864 judgements from 76 workers in \textit{Tech6$\times$6}, and 1617 judgements from 179 workers in \textit{Tech12$\times$3}.

\smallskip
\noindent\textbf{Accuracy}. The median worker accuracy resulted in $0.67$ for all of the designs and conditions, even though the distribution was elongated above the median for the conditions with highlighting.  
The Kruskal-Wallis test showed no statistically significant difference between the conditions for \textit{Tech6$\times$6} ($H(1)=0.17$, $p=.68$) and \textit{Tech12$\times$3} ($H(1)=0.20$, $p=.65$). Thus, accuracy did not improve in situations of higher cognitive demand. 

In the case of longer pages (\textit{Tech6$\times$6}), what we did observe was a huge percentage of task abandonment in the first page. The majority of workers selected the job and perhaps even tried to complete it but ultimately did not submit their contributions. 
This happened significantly more in the baseline, where only a 23\% percent of the workers assigned to the condition decided to take the task, compared to a 41\% in the condition with highlighting. 
In our experiment, this difference in task abandonment means that highlighting  can lower the perceived effort and attract more contributors, but at the same time, it also introduced a potential bias in our comparison of accuracy in attracting more committed workers in the baseline. 

\smallskip

\noindent\textbf{Decision time}. 
The median decision time for longer pages (\textit{Tech6$\times$6}) resulted in $23.43s$ for the baseline, and $13.09s$ in the highlighting condition. 
%
%
Highlighting reduced decision time by 44\% compared to no highlighting.
%
%
A Kruskal-Wallis test showed the difference between the conditions to be statistically significant ($H(1) = 36.22, p<.001$).
For longer tasks (\textit{Tech12x3}) the median decision time was of $31.45s$ in the baseline, and $20.65s$ in the highlighting condition ($H(1) = 22.80$, $p<.001$). 
%
%
In this case, highlighting reduced decision time by  34\% compared to the baseline.

\subsection{Additional analyses}

\subsubsection{Classification performance}
We computed the $F_1$ scores by aggregating the judgements by condition and highlighting quality as shown in Table \ref{table:class}, so as to assess and compare the output of the classification more in detail. 

\begin{table}[h]
\centering
\resizebox{0.9\columnwidth}{!}{%
\begin{tabular}{|l|l|l|l|l|l|l|}
\hline
 & \textbf{Bad} & \textbf{Neutral} & \textbf{Subopt} & \textbf{Good} & \textbf{All} & \textbf{None}\\ \hline
\textbf{0\%} & .333  & .700 & - & - & - & -  \\ \hline
\textbf{33\%} & .178  & .685 & .589 & .731 & - & -  \\ \hline
\textbf{66\%} & .551  & .681 & .556 & .725 & - & -  \\ \hline
\textbf{100\%} & -  & - & .71 & .744 & - & -  \\ \hline
\textbf{aggr} & - & - & - & - & .727 & -  \\ \hline
\textbf{base} & -  & - & - & - & - & .717  \\ \hline
\end{tabular}%
}
\caption{Aggregated $F_1$ scores by condition and highlighting quality for \textit{SLR-Tech} }
\label{table:class}
\end{table}

The results show that $F_1$ scores for highlighting aggregation ($F_1$: \textit{SLR-OA}=$.845$, \textit{SLR-Tech}=$.727$, \textit{Amazon}=$.945$) and ``good" highlighting in the 66\% ($F_1$: \textit{SLR-OA}=$.859$, \textit{SLR-Tech}=$.725$, \textit{Amazon}=$.953$) and 100\% conditions ($F_1$: \textit{SLR-OA}=$.845$, \textit{SLR-Tech}=$.744$, \textit{Amazon}=$.960$) to be superior to that of the baseline ($F_1$: \textit{SLR-OA}=$.830$, \textit{SLR-Tech}=$.717$, \textit{Amazon}=$.937$) for all datasets. 

This suggests that by focusing on the highlighting of highest quality, the resulting classification can be superior to that of the baseline. Interestingly, \textit{aggregating} the highlights can also result in superior classification performance, which opens up opportunities for bypassing quality annotation steps in the case of crowdsourced highlights, or using ensembles in the case of machine-generated ones. 
Part of the reason here is that useful highlighting as generated with the method described earlier (that is, by multiple independent annotators) outnumber non-useful ones, and aggregation enables to filter out the ``noise" generated by low-quality, but more rare highlights.

\subsubsection{Factors contributing to decision time and accuracy}
We performed additional analyses to investigate how the key factors of our dimensions, such as quality and length of the highlighting, document size, worker experience (meaning number of ``pages" contributed by the worker at the moment of providing the judgment), modified the impact of highlighting. 
We performed i) logistic regression analyses to predict \textit{correct judgment} (true / false) and ii) multiple regression analyses to predict \textit{decision time}, and compared the results for the baseline condition and the highlighting conditions. 
We include the regression analyses tables as supplementary materials.  
Below we summarise the main findings: 


\textit{Experience with the task increases the benefits of highlighting}. Experience (progression through the pages of the task) was a significant predictor of decision time, contributing to lower decision time in all three datasets in the highlighting conditions. For the baseline condition, it was significant for \textit{SLR-OA} and \textit{Amazon} datasets.
However, in the highlighting conditions, experience also translated in workers being less likely to make mistakes, as it was a significant predictor of correct judgment, but not in the baseline.
This insight suggests that experience, and possibly the amount of work given to the worker, increases the benefits of highlighting.

\textit{Workers adapt their behavior in longer documents}. 
%
%
The size of the document was a significant predictor of decision time and correct judgements for all datasets in the highlighting models and baseline. 
%
%
The general insight is that workers are more likely to spend more time deciding on longer documents as well as more prone to make mistakes. However, we observed that, first, this is not the case in \textit{SLR-Tech} where \textit{``long" documents predict less time to judge compared to ``short" documents in the highlighting conditions}.
Second, judging a ``long" document, despite being significant, predicts only from 1-5 seconds more in decision time than short documents, even when the length of the document is more than three times longer. Finally, deciding on ``medium" documents but not on ``long" documents increase the likelihood of incorrect judgements.

These results, and the \textit{length of the highlighting} as a significant predictor, suggest that people adapt their behavior in longer documents, possibly relying more on the highlighted text, and therefore modifying the effect of highlighting.



\subsection{Experiment 3: Impact of machine-generated highlighting}




We collected a total of $8129$ judgements from $1035$ workers. The quality distribution of the highlights generated by the automated approaches - according to the qualitative assessment - is shown in Table  \ref{table:quality-autohighlighting} to put the results into context.



\begin{table}[h]
\centering
\resizebox{.6\columnwidth}{!}{%
\begin{tabular}{llll}
\toprule
 & BertSum & Refresh & Bert-QA  \\ \midrule
\textit{SLR-OA} & .38 & .24 & .49  \\ 
\textit{SLR-Tech} & .43 & .18 & .40 \\
\textit{Amazon} & .56  & .62 & .59 \\ 
\bottomrule
\end{tabular}%
}

\caption{Proportion of useful highlights generated.}
\label{table:quality-autohighlighting}
\end{table}

\subsubsection{Worker accuracy}



\begin{figure}[h]
  \centering
  \includegraphics[width=1\columnwidth]{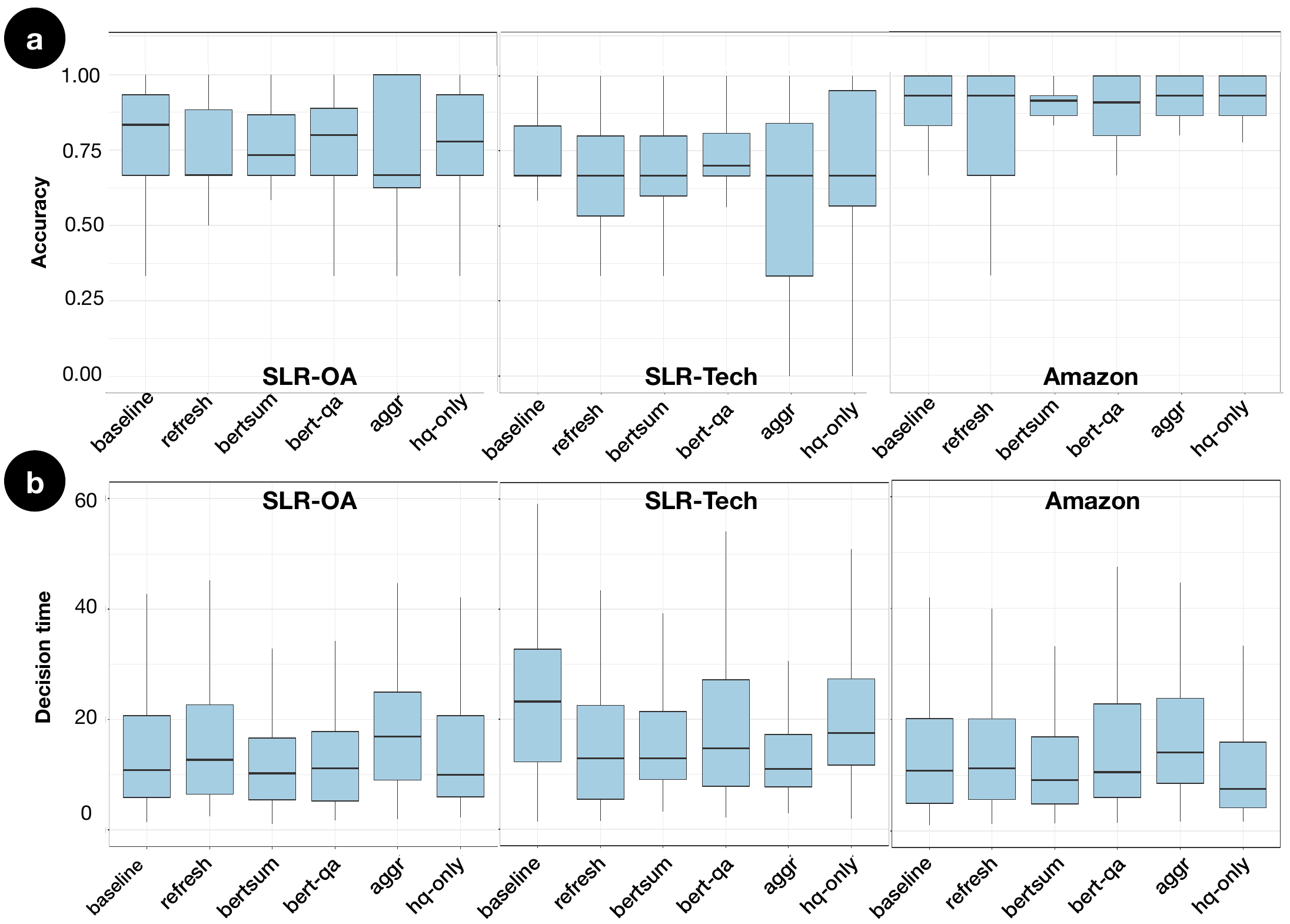}  
  \caption{Worker accuracy per condition.}~\label{fig:accuracy-ml}
\end{figure}

The conditions with highlight support did not significantly improve on accuracy over the baseline for any of the datasets. We should note, however, that the overall trend correspond to the quality distribution of each condition, i.e., lower quality translates into a lower median accuracy or elongated tail, stressing our observation that bad highlighting affects accuracy (see Figure \ref{fig:accuracy-ml}a).

\subsubsection{Decision time}
Highlighting support did not improve over the baseline for SLR-OA, which is in lines with our previous results for this dataset (Experiment 1), where highlighting support did not improve decision time regardless of the quality. In contrast, highlighting support did improve over the baseline for all conditions in SLR-Tech, as it was the case again in the first experiment. In the Amazon dataset, only the 100ML condition with high-quality highlighting improved over the baseline, but not the other conditions which are not in the promising zone (66\%-100\%) identified in the first experiment.  
The results are summarised in Figure \ref{fig:accuracy-ml}b.

\subsubsection{Classification performance}
We computed the $F_1$ scores for the aggregated performance of each condition and dataset, as shown in Table \ref{table:ml-f1}. 
%
%
The low quality of the highlighting resulted in the automated approaches performing below the baseline. Improvement, in this context, was only achieved through aggregation or selecting the best highlights among the ones available. Notice that the quality of the underlying algorithms, and the space for improvement, limits the benefit of aggregation.

\begin{table}[h]
\centering
\resizebox{1\columnwidth}{!}{%
\begin{tabular}{lllllll}
\toprule
 & BertSum & Refresh & Bert-QA & AggrML & 100ML & Base \\ \midrule
\textit{SLR-OA} & .831 & .817 & .842 & \textbf{.860} & \textbf{.863} & .858 \\ 
\textit{SLR-Tech} & .684 & .677 & .678 & .685 & .712 & .733\\ 
\textit{Amazon} & .891 & .907 & .911 & .918 & .924 & .938\\ 
\bottomrule
\end{tabular}%
}
\caption{Aggregated $F_{1}$ scores by condition. Improvements over the baseline are highlighted.}
\label{table:ml-f1}
\end{table}

\section{Discussion}
The quality assessment of the machine-generated highlights provided us with insights into the nature and potential limitations of automated approaches. 

 Extractive summarisation approaches are not trained for a specific filter and therefore are prone to generate less useful highlights. \textit{BertSum}, the algorithm of this class with the overall better performance, was particularly bad targeting ``participants" (SLR-OA), but its performance improved when targeting the ``objective" of the paper (SLR-Tech). 
 
 The Q\&A-based approach, instead, generated shorter highlights specific for each dataset and resulted in overall higher quality. However, it was sensitive to how the questions were formulated, varying in the output with each attempt. 
%
%
\textit{Bert-QA} also attempted to retrieve evidence for a question even when there was none. 
For example, if the paper is not about technology for social interaction, Bert-QA will still look for excerpts associated with these concepts,  which can sometimes lead to deceiving (bad) highlights. Instead, in these cases, a counter-argument (e.g., highlighting a different focus) is desirable, or even indicating that the question is ``unanswerable" (e.g., no highlighting at all).

The impression we got working with Bert-QA is that by training it specifically on the class of problems of interest (e.g., on SLRs in general), it could be possible to achieve a high-quality result. Attempting this is in our work pipeline.


\medskip 
Besides these considerations on ML-generated highlights, the investigation into the impact of highlighting quality provided us the following main insights: 

\begin{easylist}
\ListProperties(Hide=100, Hang=true,Style*=--~)
& \textbf{Bad highlighting support can hurt accuracy, while high quality offers no significant benefits}. High quality highlighting showed a positive trend in worker accuracy, improved over conditions of lower quality, but ultimately did not significantly improve over the baseline. Even when posing workers with tasks of higher cognitive demand, worker accuracy was not significantly better when providing good quality highlighting. The opposite however was consistent across all datasets: bad highlighting can hurt accuracy. 






& \textbf{Higher quality highlighting can reduce decision time to almost a half}. We observed that highlighting quality in the 66\% to 100\% range offered significant improvements in decision time over the baseline in two of the three datasets analysed. 
In high demand scenarios, highlighting support can reduce the decision time by 44\% compared to no highlighting, while maintaining the same level of accuracy.
{In a different domain, \cite{DBLP:conf/w4a/GaurLMB16} showed a similar insight, where automatic speech recognition (ASR) could facilitate workers at transcription tasks, but only when the ASR support was good enough.}

& \textbf{Aggregating highlighting can increase overall classification performance}. The additional analyses also uncovered the potential of aggregating highlighting by independent annotators (or algorithms), which provided benefits analogous to that of aggregating votes in crowdsourced classification: while it did not improve on individual worker accuracy, the aggregated classification performance was superior to that of the baseline. Compared to other conditions with similar accuracy, this suggests that errors in aggregated highlighting might be more independent, an interesting effect that requires further exploration.

& \textbf{Highlighting can further decrease the decision time and perceived effort in high demand scenarios}. The regression analyses also suggested that in higher demand scenarios (e.g., longer documents and increasing the number of contributions requested from workers) highlighting could increase its benefits. 
%
%
We confirmed the added benefits in terms of decision time, a reduction going from 16\% up to 44\% compared to the baseline for \textit{SLR-Tech}, as well as perceived effort (lower task abandonment), but not in terms of accuracy.
%
The difference in abandonment that we observed is in line with  \cite{DBLP:conf/wsdm/HanRGSCMD19}, where the results on relevance judgements experiments show a similar ratio of submission to abandonment; and most of the workers tend to quit early, after a quick assessment of the effort for the tasks. 
\cite{DBLP:conf/hcomp/WuQ17}  observed a similar situation, where tasks with longer instructions showed a higher abandonment rate than more compact tasks.

&  \textbf{Task difficulty does not affect the impact of highlighting}. According to our results, the impact of text highlighting on decision time was not modified by task difficulty (measured as accuracy at the baseline). The relative improvement of highlighting support in the two significant cases \textit{SLR-Tech} (accuracy: $.67$) and \textit{Amazon} (accuracy: $.94$) was of 60\% with respect to their baselines, while for \textit{SLR-OA} (accuracy: $.67$) was not significant. In the case of improvements in accuracy with respect to the baseline, the results were not significant regardless of task difficulty.

\end{easylist}

\smallskip

The takeaway message is that highlighting is a promising direction for text classification support, better suitable for situations where workers are faced with longer documents or are expected to provide a large number of contributions. Highlighting approaches should however consider the negative impact of bad highlighting, and use approaches that either i) limit the recommendations of highlighting to those with high level confidence (quality), or ii) aggregate the highlighting provided by independent annotators or algorithms -- provided that the distribution of quality favors good highlighting or is at least balanced. 

{\color{blue}
}

Experiments also show that highlighting support of good quality can significantly reduce the decision time by 44\% while maintaining (but not necessarily increasing) worker accuracy.
These benefits are elevated in situations of high cognitive demand, where workers not only see an effective decrease in decision time but also experience a lower barrier to participation.
We identified the promising quality range for highlighting support, as well as the negative effects of bad highlighting, providing alternative approaches based on highlighting aggregation and quality (or confidence) level filtering. The former is a promising direction, as it can reduce the efforts in quality annotation and allow for combining the output of ensembles of algorithms. 
We provide the datasets used in this paper in the supplementary material.

\bibliographystyle{aaai}
\bibliography{references}

\end{document}